\documentclass{article}
\usepackage{amsmath,amsthm,amssymb,authblk}

\newtheorem{theorem}{Theorem}[section]
\newtheorem{corollary}[theorem]{Corollary}
\newtheorem{proposition}[theorem]{Proposition}
\newtheorem{lemma}[theorem]{Lemma}
\newtheorem{definition}[theorem]{Definition}

\newtheorem{remark}[theorem]{Remark}
\def\pf{{\bf Proof.}~ }

\begin{document}

\title{ Rank-metric LCD codes }
\author{Xiusheng Liu,\quad Hualu Liu\\
\small School of Mathematics and Physics\\
\small Hubei Polytechnic University\\
\small  Huangshi, Hubei 435003, China}
\date{}
\maketitle

\insert\footins{\footnotesize{\it Email address}:
lxs6682@163.com (Xiusheng Liu); hwlulu@aliyun.com (Hualu Liu).}

\begin{abstract} In this paper, we investigate the rank-metric codes which are proposed by Delsarte and Gabidulin to be complementary dual codes. We point out the relationship between Delsarte complementary dual codes and Gabidulin complementary dual codes. In finite field $\mathbb{F}_{q}^{m}$, we construct two classes of  Gabidulin LCD MRD codes  by self-dual basis (or almost self-dual basis) of $\mathbb{F}_{q}^{m}$ over $\mathbb{F}_{q}$. Under a suitable condition, we determine a sufficient condition for  Delsarte optimal anticodes to be LCD codes over $\mathbb{F}_{q}$.
\end{abstract}


\bf Keywords\rm ~ Delsarte LCD codes $\cdot$  Gabidulin LCD codes $\cdot$ MRD codes $\cdot$ self-dual basis

\bf Mathematic Classifications Subject\rm  ~15A03$\cdot$15A99$\cdot$15B99
\section{Introduction}
Linear complementary dual codes (which is abbreviated to LCD codes) are linear codes that meet their dual trivially. These codes were introduced by Massey in  \cite{Massey} and showed that asymptotically good LCD codes exist, and provided an optimum linear coding solution for the two-user binary adder channel. They are also used in counter measure to passive and active side channel analyses on embedded cryto-systems  \cite{Carlet1}. Guenda, Jitman and Gulliver investigated an application of LCD codes in constructing good entanglement-assisted quantum error correcting codes \cite{Guend}.

Yang and Massey in \cite{Yang} showed that a necessary and sufficient condition for a cyclic code of length $n$ over finite fields to be an LCD code is that the generator polynomial $g(x)$ is self-reciprocal and all the monic irreducible factors of $g(x)$ have the same multiplicity in $g(x)$ as in $x^n-1$. In \cite{Sendrier}, Sendrier indicated that LCD codes  meet the asymptotic Gilbert-Varshamov bound. Esmaeili and  Yari in \cite{Esmaeili} studied LCD quasi-cyclic codes. Necessary and sufficient conditions for certain classes of quasi-cyclic codes to be LCD codes were obtained \cite{Esmaeili}.  Dougherty et al. developed a linear programming bound on the largest size of an LCD code of given length and minimum distance \cite{Dougherty}. The parameters of several classes of LCD BCH codes were explicitly determined  in \cite{Li1},\cite{Li2},\cite{Li3}.  In addition, Boonniyoma and Jitman gave a study on linear codes with Hermitian complementary dual \cite{Boonniyoma}, and we also in \cite{Liu} studied LCD codes over finite chain rings.  In recently, Jin \cite{Jin} constructed several classes of LCD MDS codes by using two classes of disjoint GRS codes. The existence question about  LCD  MDS codes over a finite field of even characteristic has been completely addressed in \cite{Jin}. Base on \cite{Jin}, Carlet et al. \cite{Carlet2} completely addressed   LCD  MDS codes over any finite field. Some other constructions of LCD MDS  codes are known (See \cite{Chen},\cite{Mes},\cite{Sar},\cite{Zhu}).

Rank-metric codes were first introduced in coding theory by Delsarte in \cite{Dels}. They are sets of matrices of fixed size, endowed with the rank distance. Rank-metric codes have cryptography applications  and applications in tape recording. Recently it was shown how to employ them for error correct in coherent linear network coding(\cite{Kot},\cite{Sil1},\cite{Sil2}). Due to these applications, there is a steady stream of work that focuses on general properties of codes with rank-metric.

An MRD code is a rank-metric code which is maximal in size given the minimum distance, in other words it achieves the Singleton bound for the rank-metric distance. Delsarte  \cite{Dels} and independently Gabidulin \cite{Gab} showed that MRD codes only exist if the size of the matrix divides the dimension of the code. More precisely, for $n\times m$ matrices with $n\leq m$, $m$ must divide the dimension of the code. According to practical applications of LCD and rank-metric codes, we study rank-metric LCD codes in this paper.

This paper is organized as follows. The necessary background materials of LCD and Delsarte rank-metric codes are given in Section 2. In Section 3,  we investigate the relationship between Delsarte LCD codes and Gabidulin LCD codes. In Section 4, we construct two classes of Gabidulin  LCD MRD codes  by self-dual basis (or almost self-dual basis) of the finite field $\mathbb{F}_{q}^{m}$ over base field $\mathbb{F}_{q}$. In Section 5, we provide a sufficient condition of  Delsarte optinal anticodes to be  LCD codes under a suitable condition,  Finally, in Section 6, a brief summary of our work is described.

\section{Preliminaries}
In this section, we recall some basic concepts and results of LCD and Delsarte rank-metric codes, necessary for the development of this work. For more details, we refer to \cite{Massey} and \cite{Dels}.

Throughout this paper, we denote by  $\mathbb{F}_{q}$ the finite  base  field  with cardinality $\mid\mathbb{F}_{q}\mid=q=p^e$ and let $\mathbb{F}_{q^m}$ be an extension field of degree $m$ with respect to base field $\mathbb{F}_{q}$, where $p$ is a prime number and $e, m $ are positive integers. It is well-known that $\mathbb{F}_{q^m}$ is isomorphic (as a vector space over $\mathbb{F}_{q}$ ) to the vector space $\mathbb{F}_{q}^m$.

We denote by $\mathbb{F}_{q}^{n\times m }$ the $\mathbb{F}_{q}$-vector space of $n\times m$ matrices with entries in $\mathbb{F}_{q}$.  The transpose of $A$  is $A^{T}$, while $\mathrm{rk}(A)$ denotes the rank of $A$. For any $A\in\mathbb{F}_{q}^{n\times m}$, we write $\mathrm{Tra}(A)$ for the trace of $A$, and $A_i=(a_{1i},a_{2i},\ldots,a_{ni})^{T}$ for the $i$-th column of $A$. The vector space generated by the columns of a matrix $A\in\mathbb{F}_{q}^{n\times m}$ is denoted by $\mathrm{colsp}(A)\subset \mathbb{F}_{q}^{n}$.

One  easily obtains the isomorphic description of matrices over the base field $\mathbb{F}_{q}$ as vector space over the extension field, i.e., $\mathbb{F}_{q}^{n\times m}\cong \mathbb{F}_{q^m}^n$.

Let $A,B\in\mathbb{F}_{q}^{n\times m}$. In \cite{Dels}, Delsarte introduced a kind of inner product,  called trace inner product, as follows:
$$
\langle A,B\rangle=\mathrm{Tra}(AB^{T}).
$$
It is easy to check that $\langle A,B\rangle=\sum_{i=1}^{m}[A_i,B_i]$, where $[,]$ denotes the Euclidean inner product on $\mathbb{F}_{q}^{n}$.

A Delsarte code of  size $n\times m$ over $\mathbb{F}_{q}$ is $\mathbb{F}_{q}$-linear subspace $C\subset\mathbb{F}_{q}^{n\times m}$. The minimum rank of a non-zero code $C$ is denoted and defined by $d_{\mathrm{r}}=\mathrm{minrk}(C)=\mathrm{min}\{\mathrm{rk}(A):~A\in C,~\mathrm{rk}(A)>0\}$, while the maximum rank of any code $C$ is  denoted and defined by $\mathrm{maxrk}(C)=\mathrm{max}\{\mathrm{rk}(A):~A\in C\}$.  The dual of $C$ is the Delsarte code
$C^{\perp}=\{B \in \mathbb{F}_{q}^{n\times m}:~\langle A,B\rangle=0,\mathrm{~for~all}~A\in C\}$.

From the fact that trace inner product is nondegenerate, it
follows immediately that $(C^{\perp})^{\perp} =C$ and $\mathrm{dim}_{\mathbb{F}_{q}}C+\mathrm{dim}_{\mathbb{F}_{q}}C^{\perp }=mn$.

The following theorem, was first proved by Delsarte,  can be found in \cite{Dels}.

\begin{theorem}\label{th: 1}$(\mathrm{\cite{Dels}, Theorem ~5.4})$ Let $C\subset\mathbb{F}_{q}^{n\times m}$ be a non-zero Delsarte code with minimum rank distance $d_{\mathrm{r}}$. Then
$$d_{\mathrm{r}}\leq\frac{mn}{max\{m,n\}}+\frac{\mathrm{dim}_{\mathbb{F}}(C)}{max\{m,n\}}+1.$$
Moreover,  for any $1\leq d_r\leq min\{m,n\}$ there exists a non-zero Delsarte code $C$ of  minimum rank distance $d_{\mathrm{r}}$ which attains the upper bound.
\end{theorem}
\begin{definition} \label{de:1}A Delsarte code $C$ is called Delsarte LCD  if $C^{\perp}\cap C=\{\mathbf{0}\} $, and a Delsarte LCD code $C$ is called Delsarte LCD  MRD if $C$ attains the  bound of $d_{\mathrm{r}}\leq\frac{mn}{max\{m,n\}}+\frac{\mathrm{dim}_{\mathbb{F}}(C)}{max\{m,n\}}+1$.
\end{definition}

\begin{remark}\label{re: 1}Note that $\mathbb{F}_{q}^{n\times m}$ is a trivial example of a Delsarte LCD MRD with minimum rank $1$ and dimension $mn$.
\end{remark}

To investigate LCD codes over $\mathbb{F}_{q^{ m}}$,  we need  the concept and a result of LCD codes.

Let $C$ be a code   over $\mathbb{F}_{q^{m}}$. If $C^{\perp}\cap C=\{\mathbf{0}\} $, Then $C$ is  called  a  LCD code.

The following theorem gives a criteria of  LCD codes and can be found in \cite{Massey}.
\begin{theorem} \label{th: 1}Let C be a linear code over $\mathbb{F} _{q^m}$ with generator matrix
$G$. Then $C$ is  LCD if and only if $GG^{T}$ is nonsingular.
\end{theorem}

\section{Delsarte and Gabidulin LCD  Codes}

In keeping with Gabidulin$^{,}$s original notation, we will use $a^{[i]}$ to mean $a^{q^{i}}$ for $a\in\mathbb{F}_{q^{m}}$ and integer $i$.

Given a vector  $(g_1,g_2,\ldots,g_n)\in \mathbb{F}_{q^m}^n$,  we denote by $M_k(g_1,g_2,\ldots,g_n)\in\mathbb{F}_{q^m}^{k\times n}$ the matrix
$$~~~~~~~~~~~~~~M_k(g_1,g_2,\ldots,g_n)=\begin{pmatrix}g_1 & g_2 & \ldots &g_n \\
     g_1^{[1]} & g_2^{[1]} & \ldots &g_n^{[1]}\\
    \vdots& \vdots & \ddots & \vdots \\
   g_{1}^{[k-1]} & g_2^{[k-1]} & \ldots &g_n^{[k-1]}\\
  \end{pmatrix}.~~~~~~~~~~~~~~~~~~~~~~~$$

A different definition of rank-metric code, proposed by Gabidulin, is the following.

 \begin{definition} \label{de: 2}$\mathrm{(\cite{Gab})}$ The rank of a vector $\alpha=(g_1,g_2,\ldots,g_n),g_i\in \mathbb{F}_{q^m}$, denoted by $\mathrm{rk}(\alpha)$, is defined as the maximal number of linearly independent coordinates $g_i$ over $\mathbb{F}_{q}$, i.e., $\mathrm{rk}(\alpha):=dim_{\mathbb{F}_q}\langle g_1,g_2,\ldots,g_n\rangle$. Then we have a metric rank  distance given by $d_{\mathrm{r}}(\alpha-\beta)=\mathrm{rk}(\alpha-\beta)$ for $\alpha,\beta\in\mathbb{F}_{q^{m}}^n $.  A Gabidulin (rank-metric) code of length $n$ with dimension $k$ over $\mathbb{F}_{q^{m}}$ is an $\mathbb{F}_{q^{m}}$-linear subspace $C\subset\mathbb{F}_{q^{m}}^n$. The minimum rank distance of a Gabidulin  code $C\neq0$ is
$$d_{\mathrm{r}}:=\mathrm{min}\{\mathrm{rk}(\alpha):~\alpha\in C,~\alpha\neq0\}.$$
\end{definition}

The Singleton bound for codes in the Hamming metric  implies also an upper bound  for Gabidulin codes.

\begin{theorem} \label{th:Gab1}$(\cite{Gab})$
Let $C\in\mathbb{F}_{q^{m}}^n$ be a Gabidulin  code with minimum rank distance $d_r$ of dimension $k$.  Then $d_r\leq n-k+1.$
\end{theorem}

A  Gabidulin  code attaining the Singleton bound  is called a Gabidulin maximum rank distance  (MRD) code.

In  his seminal paper \cite{Gab},  Gabidulin  showed the following result on MRD codes:
\begin{theorem}\label{th:Gab2}
Let $g_1,g_2,\ldots,g_n\in\mathbb{F}_{q^{m}}$ be linearly independent over $\mathbb{F}_{q}$, and let $C$ be a Gabidulin  code generated by matrix $M_k(g_1,g_2,\ldots,g_n)$. Then Gabidulin  code $C$ is a MRD code with parameters $[n,k,n-k+1]$.
\end{theorem}
The following definitions and results can be found in \cite{Rav}
\begin{definition}\label{de:3} Let $\mathcal{G}=\{a_1,a_2,\ldots,a_m\}$ be a basis of $\mathbb{F}_{q^{m}}$ over $\mathbb{F}_{q}$. The matrix associated  to a vector $\beta=(b_1,\ldots,b_n)\in\mathbb{F}_{q^{m}}^n$ with respect to $\mathcal{G}$ is the $n\times m$ matrix $M_{\mathcal{G}}(\beta)$ with entries in $\mathbb{F}_{q}$ defined by $b_i=\Sigma_{j=1}^m M_{\mathcal{G}}(\beta)_{ij}a_j$ for all $i=1,\ldots,n$. The Deslarte code associated to a Gabidulin code $C\subset\mathbb{F}_{q^{m}}^n$ with respect to the basis $\mathcal{G}$ is $\mathbf{C}_{\mathcal{G}}(C):=\{M_{\mathcal{G}}(\beta):~\beta\in C\}$.
\end{definition}

\begin{lemma} \label{le:Rav1}$(\mathrm{\cite{Rav},Proposition~15})$ Let $C$ be a Gabidulin code of length $n$ over $\mathbb{F}_{q^{m}}$. For any basis $\mathcal{G}$ of $\mathbb{F}_{q^{m}}$ over  $\mathbb{F}_{q}$,    $\mathbf{C}_{\mathcal{G}}(C)\subset \mathbb{F}_{q}^{n\times m}$ is a Delsarte code with
$$\mathrm{dim}_{\mathbb{F}_{q}}\mathbf{C}_{\mathcal{G}}(C)=m\cdot\mathrm{dim}_{\mathbb{F}_{q^m}}(C).$$
Moreover, if $C\neq0$, we have $d_r(\mathbf{C}_{\mathcal{G}}(C))=d_r(C).$
\end{lemma}
\begin{theorem} \label{th:Del1}Let $n\leq m$. Then $C\subset\mathbb{F}_{q^{m}}^n$ is a  Gabidulin  MRD code if and only if  $\mathbf{C}_{\mathcal{G}}(C)\subset \mathbb{F}_{q}^{n\times m}$ is a  Delsarte MRD code.
\end{theorem}
\pf $C\subset\mathbb{F}_{q^{m}}^n$ is a   Gabidulin MRD code if and only if $d_r(C)=n-\mathrm{dim}_{\mathbb{F}_{q^m}}(C)+1$, i.e., $md_r(C)=mn-m\mathrm{dim}_{\mathbb{F}_{q^m}}(C)+m$.

By Lemma \ref{le:Rav1}, One has $C$ is a   Gabidulin MRD code if and only if
$$d_r(C)=\frac{mn}{m}-\frac{\mathrm{dim}_{\mathbb{F}_{q}}(\mathbf{C}_{\mathcal{G}}(C))}{m}+1=\frac{mn}{\mathrm{max}\{m,n\}}-\frac{\mathrm{dim}_{\mathbb{F}_{q}}(\mathbf{C}_{\mathcal{G}}(C))}{\mathrm{max}\{m,n\}}+1,$$
 where the second equality follows from assumption $n\leq m$. Therefore, by Definition \ref{de:1}, $C\subset\mathbb{F}_{q^{m}}^n$ is a   Gabidulin MRD code if and only if  $\mathbf{C}_{\mathcal{G}}(C)\subset \mathbb{F}_{q}^{n\times m}$ is a  Delsarte MRD code.
\qed

Lemma \ref{le:Rav1} shows that any Gabidulin code can be regarded as a Delsarte
code with the same cardinality and rank distance. Clearly, since Gabidulin codes
are $\mathbb{F}_{q^m}$-linear spaces and Delsarte codes are $\mathbb{F}_q$-linear spaces, not all Delsarte
codes arise from a Gabidulin code in this way. In fact, only a few of them do. For example,
a Delsarte code $C\in \mathbb{F}_{q}^{n\times m}$ such that $dim_{\mathbb{F}_q} (C)\neq 0 ~\mathrm{mod}~m$ cannot arise from a
Gabidulin code.

In the remainder of the section we address the relationship between Delsarte LCD codes and Gabidulin LCD codes

We introduce notation and stating a few relevant preliminary results ({See\cite{Mac}).
The trace map  $\mathrm{Tr}~:~\mathbb{F}_{q^{m}}~\rightarrow~\mathbb{F}_{q}$ is defined as
$$\mathrm{Tr}(a)=a+a^{[1]}+\cdots+a^{[m-1]},~for~a\in\mathbb{F}_{q^{m}}.$$
Let $\mathcal{G}=\{a_1,a_2,\ldots,a_m\}$ be a basis of $\mathbb{F}_{q^{m}}$ over $\mathbb{F}_{q}$. Then there exists a unique basis $\mathcal{G'}=\{a_1',a_2',\ldots,a_m'\}$ such that $\mathrm{Tr}(a_ia_j')=\delta_{i,j}$, for $i,j=1,\ldots,m$, where $\delta\cdot,\cdot$ is the Kroneker delta function. $\mathcal{G'}$ is said to be the dual basis of
$\mathcal{G}$ and vice versa. $\mathcal{G}$ is said to be a self-dual basis if  $\mathcal{G}=\mathcal{G'}$.

The following results on self-dual basis is well-known.
\begin{proposition} \label{pr:1}$\mathrm{(\cite{Jung},\mathrm{Theorem}~1)}$  In  $\mathbb{F}_{q^{m}}$, a  self-dual basis  $\mathcal{G}$ exists if and only if $q$ is even or both $q$ and $m$ are odd.
\end{proposition}

\begin{lemma} \label{le:Rav2}$(\mathrm{\cite{Rav}},\mathrm{Theorem~21})$  Let $C$ be a Gabidulin code of length $n$ over $\mathbb{F}_{q^{m}}$, and let $\mathcal{G}$ be a self-dual basis of $\mathbb{F}_{q^{m}}$ over $\mathbb{F}_{q}$. Then $\mathbf{C}_{\mathcal{G}}(C^{\perp})=\mathbf{C}_{\mathcal{G}}(C)^{\perp}$.
\end{lemma}
\begin{theorem}\label{th:Del2}
Let $q$ be even or both $q$ and $m$ be odd, and let $\mathcal{G}=\{g_1,g_2,\ldots,g_m\}$ be a self-dual basis of $\mathbb{F}_{q^m}$ over  $\mathbb{F}_{q}$. Then $C\subset\mathbb{F}_{q^{m}}^n$ is a  Gabidulin LCD code if and only if  $\mathbf{C}_{\mathcal{G}}(C)\subset \mathbb{F}_{q}^{n\times m}$ is a  Delsarte LCD code. In particular, $C\subset\mathbb{F}_{q^{m}}^n$ is a   Gabidulin  LCD MRD code if and only if  $\mathbf{C}_{\mathcal{G}}(C)\subset \mathbb{F}_{q}^{n\times m}$ is a  Delsarte LCD MRD code.
\end{theorem}
\pf We firstly prove that $$\mathbf{C}_{\mathcal{G}}(C\cap C^{\perp})=\mathbf{C}_{\mathcal{G}}(C)\cap\mathbf{C}_{\mathcal{G}}(C)^{\perp}.$$

If $M_{\mathcal{G}}(\beta)\in\mathbf{C}_{\mathcal{G}}(C\cap C^{\perp})$, then $\beta\in C\cap C^{\perp}$, i.e., $\beta\in C$ and $\beta\in C^{\perp}$.

When  $\beta\in C$, we have $M_{\mathcal{G}}(\beta)\in\mathbf{C}_{\mathcal{G}}(C)$; when $\beta\in C^{\perp}$, we have $M_{\mathcal{G}}(\beta)\in\mathbf{C}_{\mathcal{G}}(C^{\perp})$. By Lemma \ref{le:Rav2}, we obtain that $M_{\mathcal{G}}(\beta)\in\mathbf{C}_{\mathcal{G}}(C)^{\perp}$. Thus, $M_{\mathcal{G}}(\beta)\in\mathbf{C}_{\mathcal{G}}(C)\cap\mathbf{C}_{\mathcal{G}}(C)^{\perp}$, i.e., $\mathbf{C}_{\mathcal{G}}(C\cap C^{\perp})\subset\mathbf{C}_{\mathcal{G}}(C)\cap\mathbf{C}_{\mathcal{G}}(C)^{\perp}.$

On the other hand, suppose that $M_{\mathcal{G}}(\beta)\in\mathbf{C}_{\mathcal{G}}(C)\cap\mathbf{C}_{\mathcal{G}}(C)^{\perp}$. Then by $M_{\mathcal{G}}(\beta)\in\mathbf{C}_{\mathcal{G}}(C)$, we have  $\beta\in C$, and by $M_{\mathcal{G}}(\beta)\in\mathbf{C}_{\mathcal{G}}(C)^{\perp}=\mathbf{C}_{\mathcal{G}}(C^{\perp})$ we have  $\beta\in C^{\perp}$. Hence $M_{\mathcal{G}}(\beta)\in\mathbf{C}_{\mathcal{G}}(C \cap C^{\perp})$,  i.e.,     $\mathbf{C}_{\mathcal{G}}(C\cap C^{\perp})\supset\mathbf{C}_{\mathcal{G}}(C)\cap\mathbf{C}_{\mathcal{G}}(C)^{\perp}.$

This prove that $$\mathbf{C}_{\mathcal{G}}(C\cap C^{\perp})=\mathbf{C}_{\mathcal{G}}(C)\cap\mathbf{C}_{\mathcal{G}}(C)^{\perp}.$$

$C\subset\mathbb{F}_{q^{m}}^n$ is a  Gabidulin LCD code if and only if $C\cap C^{\perp}=0$. Obviously,  $C\cap C^{\perp}=0$ if and only if $\mathbf{C}_{\mathcal{G}}(C\cap C^{\perp})=0$. Using above the equation,  Then $C\subset\mathbb{F}_{q^{m}}^n$ is a  Gabidulin LCD code if and only if  $\mathbf{C}_{\mathcal{G}}(C)\subset \mathbb{F}_{q}^{n\times m}$ is a  Delsarte LCD code.
\qed

The following  concept and  lemma can be found in \cite{Rav}.

\begin{definition} \label{de:3.10}Let $U\subset \mathbb{F}_{q}^n$ be a vector subspace. Then the set of matrices $A\in \mathbb{F}_{q}^{n\times m}$ with $\mathrm{colsp}(A)\subset U$ is a vector subspace of $\mathbb{F}_{q}^{n\times m}$, and we denote the vector subspace by $ \mathbb{F}_{q}^{n\times m}(U)$.
\end{definition}
\begin{lemma} \label{le:3.11}$\mathrm{(\cite{Rav},Lemma~26)}$ Let $U\subset \mathbb{F}_{q}^n$ be a vector subspace. Then $$\mathrm{dim}_{\mathbb{F}_q}(\mathbb{F}_{q}^{n\times m}(U))=m\cdot\mathrm{dim}_{\mathbb{F}_q}(U).$$
\end{lemma}
\begin{lemma} \label{le:3.12} Let $U\subset \mathbb{F}_{q}^n$ be a vector subspace. Then $\mathbb{F}_{q}^{n\times m}(U^{\perp})=(\mathbb{F}_{q}^{n\times m}(U))^{\perp}.$
\end{lemma}
\pf If $A\in\mathbb{F}_{q}^{n\times m}(U^{\perp})$, then $\mathrm{colsp}(A)\subset U^{\perp}$. This means that $[A_i,u]=0$ for any $u\in U$, where $A_i$ stand for $i$-th column of $A$, $i=1,2,\ldots,m$.

For any $B\in\mathbb{F}_{q}^{n\times m}(U)$, by $\mathrm{colsp}(B)\subset U$, we obtain $[A_i,B_i]=0$ for $i=1,2,\ldots,m$. Thus $\langle A,B\rangle=\sum_{i=1}^{m}[A_i,B_i]=0$, which implies that $$~~~~~~~~~\mathbb{F}_{q}^{n\times m}(U^{\perp})\subset(\mathbb{F}_{q}^{n\times m}(U))^{\perp}.~~~~~~~~~~~~~~~~(1)$$

On the other hand, Lemma ~\ref{le:3.11}  gives $$\mathrm{dim}_{\mathbb{F}_q}(\mathbb{F}_{q}^{n\times m}(U^{\perp}))=m\cdot \mathrm{dim}_{\mathbb{F}_q}(U^{\perp})=m(n-\mathrm{dim}_{\mathbb{F}_q}(U)),$$ and
$$\mathrm{dim}_{\mathbb{F}_q}(\mathbb{F}_{q}^{n\times m}(U))^{\perp}=m\cdot n -\mathrm{dim}_{\mathbb{F}_q}\mathbb{F}_{q}^{n\times m}(U)=m(n-\mathrm{dim}_{\mathbb{F}_q}(U)).$$
Hence,$$~~~~~~~~\mathrm{dim}_{\mathbb{F}_q}(\mathbb{F}_{q}^{n\times m}(U^{\perp}))=\mathrm{dim}_{\mathbb{F}_q}(\mathbb{F}_{q}^{n\times m}(U))^{\perp}.~~~~~~~~~~~~~(2)$$
Combining Eqs.(1) and (2) one easily obtain the lemma.
\qed
\begin{lemma} \label{le:3.13}  Let $U\subset \mathbb{F}_{q}^n$ be a vector subspace. Then $$\mathbb{F}_{q}^{n\times m}(U+U^{\perp})=\mathbb{F}_{q}^{n\times m}(U)+\mathbb{F}_{q}^{n\times m}(U^{\perp}).$$
\end{lemma}
\pf By Definition ~\ref{de:3.10}, $\mathbb{F}_{q}^{n\times m}(U+U^{\perp})=\{W:~\mathrm{colsp}(W)\subset U+U^{\perp}\}$, and $\mathbb{F}_{q}^{n\times m}(U)+\mathbb{F}_{q}^{n\times m}(U^{\perp})=\{M+N:~\mathrm{colsp}(M)\subset U,~\mathrm{colsp}(N)\subset U^{\perp}\}$.

Let $U=\langle u_1,\ldots,u_s\rangle$ and $U^{\perp}=\langle u_1',\ldots,u_t'\rangle$. For any $W\in\mathbb{F}_{q}^{n\times m}(U+U^{\perp})$, by $\mathrm{colsp}(W)\subset U+U^{\perp}$, there exist $p_{1,i},\ldots,p_{s,i},l_{1,i},\ldots,l_{t,i}\in\mathbb{F}_{q}$ such that $W_i=p_{1,i}u_1+\ldots+p_{s,i}u_s+l_{1,i}u_1'+\ldots+l_{t,i}u_t'$. Set $M_i=p_{1,i}u_1+\ldots+p_{s,i}u_s$ and $N_i=l_{1,i}u_1'+\ldots+l_{t,i}u_t'$ for $i=1,\ldots,m$. Taking $M=(M_1,\cdots,M_m)$ and $N=(N_1,\cdots,N_m)$, we have $\mathrm{colsp}(M)\subset U$ and $\mathrm{colsp}(N)\subset U^{\perp}$. Obviously, $W=M+N$. Thus, $W\in\mathbb{F}_{q}^{n\times m}(U)+\mathbb{F}_{q}^{n\times m}(U^{\perp})$, which implies that   $$~~~~~~~~~\mathbb{F}_{q}^{n\times m}(U+U^{\perp})\subset\mathbb{F}_{q}^{n\times m}(U)+\mathbb{F}_{q}^{n\times m}(U^{\perp}).~~~~~~~~~~~~~~~(3)$$

Conversely, if $Q\in \mathbb{F}_{q}^{n\times m}(U)+\mathbb{F}_{q}^{n\times m}(U^{\perp})$, then there exist $X,Y\in \mathbb{F}_{q}^{n\times m}$ such that  $\mathrm{colsp}(X)\subset U$ , $\mathrm{colsp}(Y)\subset U^{\perp}$, and $Q=X+Y$. It is easy to see that $\mathrm{colsp}(X+Y)\subset \mathrm{colsp}(X)+\mathrm{colsp}(Y)$. Thus, we have $\mathrm{colsp}(X+Y)\subset (U+U^{\perp})$, i.e., $Q\in\mathbb{F}_{q}^{n\times m}(U+U^{\perp})$. This means that $$~~~~~~~~\mathbb{F}_{q}^{n\times m}(U+U^{\perp})\supset\mathbb{F}_{q}^{n\times m}(U)+\mathbb{F}_{q}^{n\times m}(U^{\perp}).~~~~~~~~~~~~(4)$$
Combining Eqs.(3) and (4) one easily obtain the lemma.
\qed

According to the concept and the lemmas above, we can prove the following theorem.
\begin{theorem}\label{th:3.14}
Let $U\subset \mathbb{F}_{q}^n$ be a vector subspace. Then we have
$$ ~~~~~~~~~~\mathbb{F}_{q}^{n\times m}(U)\cap \mathbb{F}_{q}^{n\times m}(U^{\perp})= \mathbb{F}_{q}^{n\times m}(U\cap U^{\perp}).~~~~~~~~~~~~~~~~(5)$$
Further,  $U$ is LCD if and only if $ \mathbb{F}_{q}^{n\times m}(U)$ is LCD.
\end{theorem}
\pf  Assume that $A\in \mathbb{F}_{q}^{n\times m}(U)\cap \mathbb{F}_{q}^{n\times m}(U^{\perp})$. Then $\mathrm{colsp}(A)\subset U$ and $\mathrm{colsp}(A)\subset U^{\perp}$, i.e., $\mathrm{colsp}(A)\subset U \cap U^{\perp}$. Therefore, $$~~~~~~~~~~~ \mathbb{F}_{q}^{n\times m}(U)\cap \mathbb{F}_{q}^{n\times m}(U^{\perp})\subset\mathbb{F}_{q}^{n\times m}(U\cap U^{\perp}).~~~~~~~~~~~~~~~(6)$$

On the other hand, by Lemma~\ref{le:3.11} and ~\ref{le:3.13}, we have
$$\mathrm{dim}_{\mathbb{F}_q}(\mathbb{F}_{q}^{n\times m}(U)\cap \mathbb{F}_{q}^{n\times m}(U^{\perp}))=\mathrm{dim}_{\mathbb{F}_q}(\mathbb{F}_{q}^{n\times m}(U))+\mathrm{dim}_{\mathbb{F}_q}(\mathbb{F}_{q}^{n\times m}(U^{\perp}))$$
$$~~~~~~~~~~~~~~~~~~~~~~~~~~~~~~~~~~~~~~~~~~~~~~-\mathrm{dim}_{\mathbb{F}_q}(\mathbb{F}_{q}^{n\times m}(U)+\mathbb{F}_{q}^{n\times m}(U^{\perp}))$$
$$~~~~~~~~~~~~~~~~~~~~~~~~~~~~~~~~~~~~~~~~~~~=m\cdot n-\mathrm{dim}_{\mathbb{F}_q}(\mathbb{F}_{q}^{n\times m}(U+U^{\perp}))$$
$$~~~~~~~~~~~~~~~~~~~~~~~~~~~~~~~~~~~~~~~~=m\cdot(n-\mathrm{dim}_{\mathbb{F}_q}(U+ U^{\perp})),~~~~~~~~~~~~~~~~~~~~~~~~~~~~~~~$$
and
$$\mathrm{dim}_{\mathbb{F}_q}(\mathbb{F}_{q}^{n\times m}(U\cap U^{\perp}))=m\cdot\mathrm{dim}_{\mathbb{F}_q}(U\cap U^{\perp})~~~~~~~~~~~~~~~~~$$
$$~~~~~~~~~~~~~~~~~~~~~~~~~~~=m\cdot(\mathrm{dim}_{\mathbb{F}_q}(U)+\mathrm{dim}_{\mathbb{F}_q}(U^{\perp})-\mathrm{dim}_{\mathbb{F}_q}(U+ U^{\perp}))$$
$$~~~~~~~~~~~~~~~~~~~~~~~~~~~~~~~~~~~~~~~~=m\cdot(n-\mathrm{dim}_{\mathbb{F}_q}(U+ U^{\perp})).~~~~~~~~~~~~~~~~~~~~~~~~~~~~~~~$$
Thus,
$$~~~~~~\mathrm{dim}_{\mathbb{F}_q}(\mathbb{F}_{q}^{n\times m}(U)\cap \mathbb{F}_{q}^{n\times m}(U^{\perp}))=\mathrm{dim}_{\mathbb{F}_q}(\mathbb{F}_{q}^{n\times m}(U\cap U^{\perp})).~~~~~~~~~~(7)$$
Combining Eqs.(6) and (7) one  obtain the equation $\mathbb{F}_{q}^{n\times m}(U)\cap \mathbb{F}_{q}^{n\times m}(U^{\perp})= \mathbb{F}_{q}^{n\times m}(U\cap U^{\perp})$.

By Lemma \ref{le:3.12}, the second statement is obvious.
\qed

\section{ Gabidulin LCD MRD codes}
In this section we will give two classes of Gabidulin LCD  MRD codes by almost self-dual basis of $\mathbb{F}_{q}^{m}$ over $\mathbb{F}_{q}$. We first introduce  the definition of almost self-dual basis and the theorem for the existence of almost self-dual basis(See\cite{Jung}).
\begin{definition} The elements $g_1,g_2,\ldots,g_m\in\mathbb{F}_{q^m}$ form an almost self-dual basis of $\mathbb{F}_{q^m}$ over $\mathbb{F}_{q}$ if and only if
$$ Tr(g_ig_j)=0,~for~i\neq j,~i,j=1,\ldots,m,$$
and
$$Tr(g_1^2)=\cdots=Tr(g_{m-1}^2)=1,~Tr(g_m^2)=a\neq0,$$
where $a=1$ or $a\neq1$.
\end{definition}
\begin{theorem}\label{th:Jung} $\mathrm{(\cite{Jung},\mathrm{Theoem}~2)}$For any odd $q$, $\mathbb{F}_{q^m}$ has an almost  a self-dual basis of $\mathbb{F}_{q}$.
\end{theorem}
\begin{theorem}\label{th:3.15}
Let $q$ be even, and let $\mathcal{G}=\{g_1,g_2,\ldots,g_m\}$ be a self-dual basis of $\mathbb{F}_{q^m}$ over  $\mathbb{F}_{q}$. If $C$ is  a Gabidulin code of length $n=m$ over $\mathbb{F}_{q^m}$ generated by matrix
$$~~~~~~~~~~~~~~G=\begin{pmatrix}g_1 & g_2 & \ldots &g_m \\
     g_1^{[1]} & g_2^{[1]} & \ldots &g_m^{[1]}\\
    \vdots& \vdots & \ddots & \vdots \\
   g_{1}^{[k-1]} & g_2^{[k-1]} & \ldots &g_m^{[k-1]}\\
  \end{pmatrix},~~~~~~~~~~~~~~~~~~~~~~~$$
then $C$ is a   Gabidulin LCD MRD code with parameters $[n,k,d_r]$, where $d_r\leq n$.
\end{theorem}
\pf Since $\mathcal{G}=\{g_1,g_2,\ldots,g_m\}$  is a self-dual basis of  over  $\mathbb{F}_{q}$,  we have $\mathrm{Tr}(g_ig_j)=\delta_{ij}$ for $i,j=1,2,\ldots,m$. Taking
$$A=\begin{pmatrix}g_1 & g_1^{[1]} & \ldots &g_1^{[m-1]} \\
     g_2 & g_2^{[1]} & \ldots &g_2^{[m-1]}\\
    \vdots& \vdots & \ddots & \vdots \\
   g_{m} & g_m^{[1]} & \ldots &g_m^{[m-1]}\\
  \end{pmatrix}.$$
Thus, $AA^T=I_m$.  Then we also have $$A^TA=\begin{pmatrix}\sum_{i=1}^mg_i^2 &\sum_{i=1}^mg_i^{1+q} & \ldots &\sum_{i=1}^mg_i^{1+q^{m-1}}\\
    \sum_{i=1}^mg_i^{1+q} & \sum_{i=1}^mg_i^{2q} & \ldots &\sum_{i=1}^mg_i^{q+q^{m-1}}\\
    \vdots& \vdots & \ddots & \vdots \\
   \sum_{i=1}^mg_i^{1+q^{m-1}} & \sum_{i=1}^mg_i^{q+q^{m-1}} & \ldots &\sum_{i=1}^mg_i^{2q^{m-1}}\\ 
  \end{pmatrix}=I_m.$$
It follows that
$$~~~~~~~~~~~~~~GG^T=\begin{pmatrix}g_1 & g_2 & \ldots &g_m \\
     g_1^{[1]} & g_2^{[1]} & \ldots &g_m^{[1]}\\
    \vdots& \vdots & \ddots & \vdots \\
   g_{1}^{[k-1]} & g_2^{[k-1]} & \ldots &g_m^{[k-1]}\\ 
  \end{pmatrix}\cdot\begin{pmatrix}g_1 & g_1^{[1]} & \ldots &g_1^{[k-1]} \\
  g_2 & g_2^q & \ldots &g_2^{[k-1]}\\
    \vdots& \vdots & \ddots & \vdots  \\
   g_{m} & g_m^{[1]} & \ldots &g_m^{[k-1]}\\ \end{pmatrix}=I_k.$$

From Theorem \ref{th: 1}  and \ref{th:Gab2}, therefore, $C$ is a  Gabidulin LCD MRD code with parameters $[n,k,d_r]$.
\qed

\begin{theorem}\label{th:A}
Let $q$ be odd, and let $\mathcal{G}=\{g_1,g_2,\ldots,g_m\}$ be an almost self-dual basis of $\mathbb{F}_{q^m}$ over  $\mathbb{F}_{q}$. If $C$ is  a Gabidulin code of length $n=m$ over $\mathbb{F}_{q^m}$ generated by matrix
$$~~~~~~~~~~~~~~G=\begin{pmatrix}g_1 & g_2 & \ldots &g_m \\
     g_1^{[1]} & g_2^{[1]} & \ldots &g_m^{[1]}\\
    \vdots& \vdots & \ddots & \vdots \\
   g_{1}^{[k-1]} & g_2^{[k-1]} & \ldots &g_m^{[k-1]}\\ 
  \end{pmatrix},~~~~~~~~~~~~~~~~~~~~~~~$$
then $C$ is a   Gabidulin LCD MRD code with parameters $[n,k,d_r]$, where $d_r\leq n$.
\end{theorem}
\pf Since $\mathcal{G}=\{g_1,g_2,\ldots,g_m\}$  is an almost self-dual basis of  over  $\mathbb{F}_{q}$,  we have $\mathrm{Tr}(g_ig_j)=\delta_{ij}$ for $i,j=1,2,\ldots,m$. Taking
$$A=\begin{pmatrix}g_1 & g_1^{[1]} & \ldots &g_1^{[m-1]} \\
     g_2 & g_2^{[1]} & \ldots &g_2^{[m-1]}\\
    \vdots& \vdots & \ddots & \vdots \\
   g_{m} & g_m^{[1]} & \ldots &g_m^{[m-1]}\\ 
  \end{pmatrix}.$$
Thus, $$AA^T=\begin{pmatrix}1 &0& \ldots &0&0\\
   0& 1 & \ldots &0&0\\
    \vdots& \vdots & \ddots & \vdots& \vdots\\
    0&0&\ldots&1&0\\
   0&0&\ldots&0&a\\
  \end{pmatrix}.$$

  Then we also have $$A^TA=\begin{pmatrix}\sum_{i=1}^mg_i^2 &\sum_{i=1}^mg_i^{1+q} & \ldots &\sum_{i=1}^mg_i^{1+q^{m-1}}\\
    \sum_{i=1}^mg_i^{1+q} & \sum_{i=1}^mg_i^{2q} & \ldots &\sum_{i=1}^mg_i^{q+q^{m-1}}\\
    \vdots& \vdots & \ddots & \vdots \\
   \sum_{i=1}^mg_i^{1+q^{m-1}} & \sum_{i=1}^mg_i^{q+q^{m-1}} & \ldots &\sum_{i=1}^mg_i^{2q^{m-1}}\\ 
  \end{pmatrix}=\begin{pmatrix}1 &0& \ldots &0&0\\
   0& 1 & \ldots &0&0\\
    \vdots& \vdots & \ddots & \vdots& \vdots\\
    0&0&\ldots&1&0\\
   0&0&\ldots&0&a\\ 
  \end{pmatrix}.$$
It follows that, when $k<m$,
$$~~~~~~~~~~~~~~GG^T=\begin{pmatrix}g_1 & g_2 & \ldots &g_m \\
     g_1^{[1]} & g_2^{[1]} & \ldots &g_m^{[1]}\\
    \vdots& \vdots & \ddots & \vdots \\
   g_{1}^{[k-1]} & g_2^{[k-1]} & \ldots &g_m^{[k-1]}\\ 
  \end{pmatrix}\cdot\begin{pmatrix}g_1 & g_1^{[1]} & \ldots &g_1^{[k-1]} \\
  g_2 & g_2^q & \ldots &g_2^{[k-1]}\\
    \vdots& \vdots & \ddots & \vdots  \\
   g_{m} & g_m^{[1]} & \ldots &g_m^{[k-1]}\\ \end{pmatrix}=I_k;$$
when $k=m$,
$$GG^T=\begin{pmatrix}g_1 & g_2 & \ldots &g_m \\
     g_1^{[1]} & g_2^{[1]} & \ldots &g_m^{[1]}\\
    \vdots& \vdots & \ddots & \vdots \\
   g_{1}^{[k-1]} & g_2^{[k-1]} & \ldots &g_m^{[k-1]}\\ 
  \end{pmatrix}\cdot\begin{pmatrix}g_1 & g_1^{[1]} & \ldots &g_1^{[k-1]} \\
  g_2 & g_2^q & \ldots &g_2^{[k-1]}\\
    \vdots& \vdots & \ddots & \vdots  \\
   g_{m} & g_m^{[1]} & \ldots &g_m^{[k-1]}\\ \end{pmatrix}=\begin{pmatrix}1 &0& \ldots &0&0\\
   0& 1 & \ldots &0&0\\
    \vdots& \vdots & \ddots & \vdots& \vdots\\
    0&0&\ldots&1&0\\
   0&0&\ldots&0&a\\ 
  \end{pmatrix}.$$

From Theorem \ref{th: 1}  and \ref{th:Gab2}, therefore, $C$ is a Gabidulin LCD  MRD code with parameters $[n,k,d_r]$.
\qed

Let $C$ be an $[n,k,d_r]$ MRD code of length $n$ over  $\mathbb{F}_{q^m}$ ($n\leq m$), and $C^{s}:=C\times\cdots\times C $ be the code obtained by $s$ cartesian products of $C$. Then $C^s$ is  code with length $n'=ns$, dimension $k'=ks$, and minimum rank distance $d'_r=d_r=n-k+1$. It can be proved that $C^s$ is an MRD codes if and only if $n=m$ (See\cite{GabM}). Note that  Gabidulin  MRD  codes have length $n\leq m$, whereas the $s$ cartesian products of  MRD codes $C$ have length $n'=sn=sm\geq m$.

In order to investigate $C^s$ to be  LCD, we need the following lemma.

\begin {lemma}\label{le:3.16} Let $C$ be a linear code of length $n$ over  $\mathbb{F}_{q^m}$. Then $(C^s)^{\perp}=(C^{\perp})^s$.
\end {lemma}
\pf Assume that $\alpha=(a_1,\ldots,a_s)\in (C^{\perp})^{s}$. Then, for $i=1,\ldots,s$, $a_i\in C^\perp$, i.e., $[a_i,b]=0$ for any $b\in C$. Thus, for any $\beta=(b_1,\ldots,b_s)\in C^s$, we have $$[\alpha,\beta]=\Sigma_{i=1}^s[a_i,b_i]=0,$$
which implies that $\alpha \in (C^s)^{\perp}$, i.e., $(C^{\perp})^s\subset(C^s)^{\perp}$.

Conversely, assume that $\omega=(w_1,\ldots,w_s)\in (C^{s})^{\perp}$. Then, for any $\nu=(v_1,\ldots,v_s)\in C^{s}$, we have $$~~~~~~~~~~~~[\omega,\nu]=\Sigma_{i=1}^s[w_i,v_i]=0.~~~~~~~~~~~~~~~~~~~~~(8)$$ 
We assume that $v_0=v_{s+1}=0$. For $i=1,\ldots,s$, taking $v_0=\ldots=v_{i-1}=v_{i+1}=\ldots=v_{s+1}=0$ in above Eqs.(8), we get $[\omega,\nu]=[w_i,v_i]=0$ for $i=1,\ldots,s$. This means that $w_i\in C^{\perp}$ for $i=1,\ldots,s$, i.e.,  $\omega\in (C^{\perp})^s$. Thus, $(C^{\perp})^s\supset(C^s)^{\perp}$, which proves the lemma.
\qed

\begin{theorem}\label{th:3.17}
$C^s$ is a    LCD MRD code of length $n'=sn$ over $\mathbb{F}_{q^m}$ if and only if $C$ is a LCD MRD code of length $n=m$ over $\mathbb{F}_{q^m}$.
\end{theorem}
\pf Sufficient: Obviously, we only need to prove $C^s$ is LCD. Assume that $\alpha=(a_1,\ldots,a_s)\in C^s\cap (C^s)^{\perp}$.  Then by Lemma \ref{le:3.16}, we have $a_i\in C\cap C^\perp$ for $i=1,\ldots,s$. Since $C$ is LCD, $a_i=0$ for $i=1,\ldots,s$. Hence, $C^s \cap(C^s)^{\perp}=0$, i.e., $C^s$ is a LCD MRD code of length $n'=sn$ over $\mathbb{F}_{q^m}$.

Necessary: Assume that $b\in C\cap C^\perp$. Then $\beta=(b,\ldots,b)\in C^s\cap (C^{\perp})^s$. Since $C^s$ is LCD,  $C^s\cap (C^{\perp})^s=C^s\cap (C^s)^{\perp}=0$ by Lemma \ref{le:3.16}. Thus, $b=0$, i.e., $C$ is a  LCD MRD code of length $n$ over $\mathbb{F}_{q^m}$.
\qed

By Theorem \ref{th:Del1}, \ref{th:Del2}, \ref{th:3.15},~\ref{th:A} and \ref{th:3.17}, we immediately obtain the following corollary.

\begin{corollary}\label{co:4.17} Let $\mathcal{G}=\{g_1,g_2,\ldots,g_m\}$ be a self-dual basis or an almost of $\mathbb{F}_{q^m}$ over  $\mathbb{F}_{q}$. If $C$ is  a Gabidulin code of length $n=m$ over $\mathbb{F}_{q^m}$ generated by matrix
$$~~~~~~~~~~~~~~G=\begin{pmatrix}g_1 & g_2 & \ldots &g_m \\
     g_1^{[1]} & g_2^{[1]} & \ldots &g_m^{[1]}\\
    \vdots& \vdots & \ddots & \vdots \\
   g_{1}^{[k-1]} & g_2^{[k-1]} & \ldots &g_m^{[k-1]}\\
  \end{pmatrix},~~~~~~~~~~~~~~~~~~~~~~~$$
then $\mathbf{C}_{\mathcal{G}}(C)$ and $\mathbf{C}_{\mathcal{G}}(C)^{s}$ are Delsarte LCD MRD codes over $\mathbb{F}_{q}$.
\end{corollary}

\section{Delsarte LCD optimal anticodes}
In this section, we provide a sufficient condition for  Delsarte optimal anticodes to be LCD codes under a suitable condition. Let us first briefly recall the definition and results for Delsarte optinal anticodes (See \cite{Rav}).
\begin{proposition}\label{pro:5.1}$(\cite{Rav},~\mathrm{Proposition ~47})$ Let $C\subset\mathbb{F}_{q}^{n\times m}$  be a Delsarte code. Then
$$\mathrm{dim}_{\mathbb{F}_{q}}(C)\leq \mathrm{max}\{n,m\}\cdot \mathrm{maxrk}(C).$$
Moreover, for any choice of $0\leq l \leq \mathrm{min}\{n,m\}$ there exists a code $C\subset\mathbb{F}_{q}^{n\times m}$ with maximum rank equal to $l$ and attaining the upper bound.
\end{proposition}
According to proposition above, we introduce the following definition.
\begin{definition}\label{def:5.2}$(\cite{Rav},~\mathrm{Deitiofinition ~48})$ A code  $C\subset\mathbb{F}_{q}^{n\times m}$ which attains the upper bound $\mathrm{dim}_{\mathbb{F}_{q}}(C)\leq \mathrm{max}\{n,m\}\cdot \mathrm{maxrk}(C)$ is said to be a Delsarte optimal abticode. Moreover, a Delsarte optimal anticode $C$ is said to be a Delsarte LCD optimal anticode, if $C$ is a LCD code.
\end{definition}
\begin{theorem}\label{th:5.3}$(\cite{Rav},~\mathrm{Theorem ~54})$ If $C$ is  a Delsarte  optimal anticode, then  $C^{\perp}$ is also a Delsarte  optimal anticode.
\end{theorem}
\begin{theorem}\label{th:5.4} Let $C$ be  a Delsarte  optimal anticode. If  $$\mathrm{minrk}(C)>\mathrm{min}\{n,m\}-\mathrm{maxrk}(C),$$ then  $C$ is  a Delsarte  LCD optimal anticode.
\end{theorem}
\pf  Since $C$ is  a Delsarte  optimal anticode,  we know that  $C^{\perp}$ is also a Delsarte  optimal anticode by Theorem \ref{th:5.3}. According to Definition \ref{def:5.2},  we have $\mathrm{dim}_{\mathbb{F}_{q}}(C)= \mathrm{max}\{n,m\}\cdot \mathrm{maxrk}(C)$ and $\mathrm{dim}_{\mathbb{F}_{q}}(C^{\perp})= \mathrm{max} \{n,m\}\cdot \mathrm{maxrk}(C^{\perp})$.

On the other hand, since $\mathrm{dim}_{\mathbb{F}_{q}}(C)+\mathrm{dim}_{\mathbb{F}_{q}}(C^{\perp})=nm$, we have $\mathrm{dim}_{\mathbb{F}_{q}}(C^{\perp})=nm-\mathrm{max}\{n,m\}\cdot \mathrm{maxrk}(C)$. Thus $\mathrm{maxrk}(C^{\perp})=\mathrm{min}\{n,m\}-\mathrm{maxrk}(C)$. We clear have $C\cap C^{\perp}=\{0\}$, i.e., $C$ is  a Delsarte  LCD optimal anticode.
\qed

\section{Conclusion}
First, we address the relationship between Delsarte LCD codes and Gabidulin LCD codes. Then we give two methods of constructing   Gabidulin LCD MRD codes by self-dual basis (or almost self-dual basis) of the finite field $\mathbb{F}_{q}^{m}$ over base field $\mathbb{F}_{q}$. Finally, we obtain a sufficient condition for  Delsarte optimal anticodes to be LCD codes under a suitable condition. In a future work, we will discuss the conditions for generalized Gabidulin MRD codes to be LCD codes.

\section*{Acknowledgment}
This work was supported by Research Funds of Hubei Province, Grant
No. D20144401,  the Educational Commission of Hubei Province, Grant No. B2015096, and Research Project of Hubei Polytechnic University, Grant No. 17xjzo3A.

\end{document}